# Reinforcement Learning for Protocol Synthesis in Resource-Constrained Wireless Sensor and IoT Networks

Hrishikesh Dutta, Amit Kumar Bhuyan, and Subir Biswas
*Michigan State University, East Lansing, USA*
duttahr1@msu.edu, bhuyanam@msu.edu, sbiswas@egr.msu.edu

*Abstract* — This article explores the concepts of online protocol synthesis using Reinforcement Learning (RL). The study is performed in the context of sensor and IoT networks with ultra-low-complexity wireless transceivers. The paper introduces the use of RL and Multi Arm Bandit (MAB), a specific type of RL, for Medium Access Control (MAC) under different network and traffic conditions. It then introduces a novel learning-based protocol synthesis framework that addresses specific difficulties and limitations in medium access for both random access and time-slotted networks. The mechanism does not rely on carrier-sensing, network time-synchronization, collision detection, and other low-level complex operations, thus making it ideal for ultra-simple transceiver hardware used in resource-constrained sensor and IoT networks. Additionally, the ability of independent protocol learning by the nodes makes the system robust and adaptive to the changes in network and traffic conditions. It is shown that the nodes can be trained to learn to avoid collisions, and to achieve network throughputs that are comparable to ALOHA-based access protocols in sensor and IoT networks with simplest transceiver hardware. It is also shown that using RL, it is feasible to synthesize access protocols that can sustain network throughput at high traffic loads, which is not feasible in the ALOHA-based systems. The system's ability to provide throughput fairness under network and traffic heterogeneities are also experimentally demonstrated.

## I. INTRODUCTION

Traditionally, wireless network protocols are designed based on heuristics and past experience of human designers. Most of the well-known wireless access protocols such as ALOHA, CSMA, and their derivatives including Bluetooth, Zigbee, and WiFi are products of such design processes [1, 2]. The choice of a network protocol is often steered by the availability of transceiver level hardware support for carrier sensing, collision detection, communication energy constraints, etc. In spite of their general success, these approaches do underperform under certain topology and traffic load heterogeneities, and specialized prioritization requirements. For instance, in case of the well-known ALOHA and SLOTTED-ALOHA MAC logics, a surge in network traffic can lead to a complete throughput collapse caused by collision avalanches. Such phenomena are particularly harmful for IoT and Sensor networks in which energy and other resource wastage can be operationally detrimental. Such effects are aggravated for heterogeneous traffic and topological diversities. Furthermore, topologically disadvantageous nodes in an arbitrary mesh network may not receive a fair share of bandwidth due to its disproportionate collision experience. All these effects point to a need for alternative protocol design approaches beyond the existing empirical designs.

To that end, Reinforcement Learning (RL) has been applied in the literature [3-14] for protocol synthesis via online learning. A protocol constitutes inter-node transmission logic, which is modeled as a Multi-Agent Markov Decision Process (MA-MDP) problem. Such MA-MDPs are then solved using an online temporal difference solution approach, namely RL. The online learning ability of RL makes the nodes learn and adapt to the best transmission logic (i.e., protocol) on the fly without *a priori* training. Additionally, the multi-agent approach enables independent learning for the node, thus making the solutions more robust and adaptive.

Such learning can be explored in two broad areas of MAC logics, namely, random access and scheduled with time-slotting. While the first category including ALOHA, CSMA, and their higher order derivatives can be synthesized using traditional RL [15], for scheduled access such as TDMA would need a special class of RL without state abstraction, known as Multi-Armed Bandits (MAB).

The existing work in this area has the following limitations. First, most of the RL solutions are centralized [5, 6] in which a single learning entity maintains current network-level information and learns transmission policies for all the network nodes. This entails frequent node-to-learner information and learner-to-node policy transfers, requiring additional control plane bandwidth. Moreover, the learner requires to maintain a network-scale learning table which adds to its storage and computation expenses. These bandwidth, storage, and single point of computation overheads make centralized learning non-scalable and vulnerable to single point of failure. The second major limitation is that network and traffic heterogeneities and traffic prioritization are neglected in the existing techniques [9]. This makes some of these approaches unsuitable in application-specific networks with specialized network configurations and performance needs. Additionally, many of the existing RL solutions assume non-sensor and IoT friendly complex transceiver capabilities including carrier-sensing, collision detection in few cases, and network time-synchronization for the MAB-based transmission scheduling.

This paper attempts to avoid those limitations using a novel RL and MAB-based learning approach for synthesizing MAC logic. The key approach here is to leverage interactive individual learning, where each node learns transmission policies independently by observing the impacts of their RL/MAB transmission actions on collisions experienced by all other nodes in the neighborhood. This is done without carrier-sensing, collision detection, and time-synchronization, thus making it suitable for low-complexity and resource-constrained networks. Specifically, the developed framework caters to two broad classes of medium access schemes, viz, random access and scheduling-based. It makes the nodes learn independently in order to attain and maintain the maximum achievable throughput for random access, and to obtain a collision-free slot allocation in scheduling-based approaches. Fig. 1 shows a generalized system architecture of the IoT network with the embedded learning components, where each IoT node acts as a learning agent. With the long-term goal of developing a generalized learning framework for protocol synthesis, this paper specifically demonstrates the concept of protocol synthesis in resource-constrained networks with low-complex transceivers not relying on aforementioned complex hardware requirements.

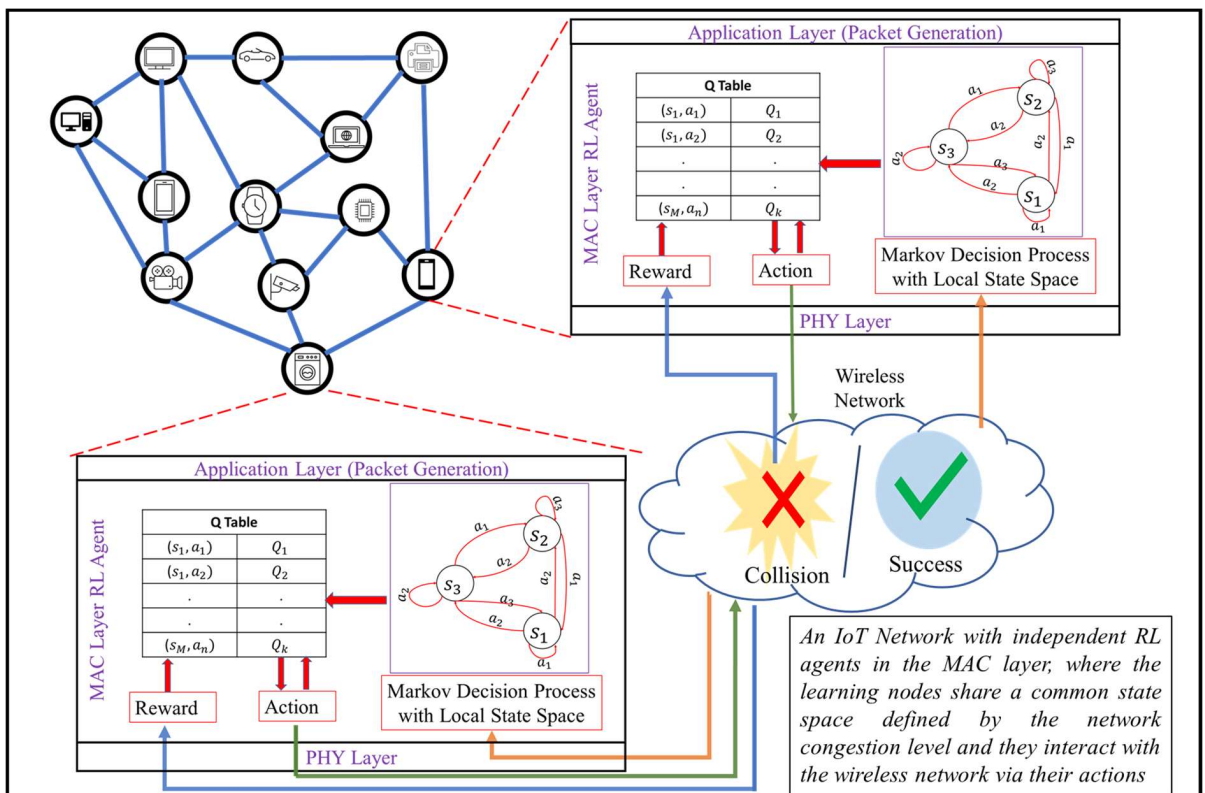

Fig. 1: System Level Architecture of an IoT Network with Embedded Learning Components

Specific contributions of this work are as follows. First, an online learning-based framework is developed for minimizing packet collisions in resource-constrained networks with random access and scheduling-based Medium Access schemes. Second, a novel slot-defragmentation mechanism is proposed for handling the trade-off between learning convergence time and spectral usage efficiency in transmission scheduling in networks without time synchronization. Third, the developed framework is decentralized such that each node learns its own transmission schedule independently relying only on localized neighborhood information. Finally, the developed learning framework is functionally validated, and performance is evaluated under heterogeneous network and traffic conditions with extensive simulation experiments.

## II. RELATED WORKS

Many Reinforcement Learning (RL) based approaches were proposed in the literature for wireless MAC protocol synthesis. The paper in [3] uses RL for wireless sensor network MAC to minimize energy expenditure while maximizing throughput. It works with slotted time and uses stateless Q-learning for nodes to find collision-free transmission slots. Q-learning-based protocols for resource allocation are also proposed in [5,10]. These mechanisms can learn and adapt with new and departing nodes while maximizing throughput. Using carrier-sensing, the nodes learn to transmit/wait [10] or to increase/decrease access contention window [5] to reduce collisions. The mechanism in [4] uses RL for solving a Partially Observable Markov Decision Process (POMDP) in order to minimize the interference amongst primary and secondary users in a cognitive network.

Researchers have also used RL and its variants for slot scheduling in TDMA-based MAC systems. An RL-based MAC protocol is proposed in [7], which improves network throughput by reducing collisions in a time-synchronous slotted network. Using stateless Q-learning nodes learn to transmit in collision-free slots. The mechanism in [8] allows nodes to learn radio schedules based on instantaneous packet traffic load in their immediate neighborhoods. The mechanism in [9] minimizes MAC layer energy expenditure via RL-based learning. Such learned low-energy protocols with sleep/active scheduling are claimed to be useful for high-density communication in wireless sensor networks. A learning-based slot allocation scheme is developed in [12] for optimizing energy and packet delay in large networks with high traffic loading. Another RL-based congestion control scheme for satellite IoT networks is proposed in [13], where the aim is to allocate channels efficiently in a TSCH network. The proposed mechanism relies on centralized arbitration at a satellite. The framework presented in [14] uses Multi-Armed Bandits (MAB) to learn an optimal back-off period in a contention-based time-slotted underwater network. The objective is to simultaneously minimize collisions and energy with the assistance of a centralized arbitration. Apart from the scalability issues of centralized RL approaches [16-18], the proposed policies require individual end nodes to download learnt policies, thus requiring additional bandwidth/channel for such control information sharing.

All these RL-based MAC frameworks rely on various combinations of underlying hardware features such as time-slotting, time-synchronization, and carrier-sensing, which can often be infeasible for ultra-resource-constrained sensor and IoT nodes. In this paper, the main focus is to explore online learning using RL and its variants for networks without such complex and energy-expensive features. The paper first demonstrates the feasibility of these learning frameworks to maximize performance in networks using random access schemes without time-slotting ability. It is shown how the maximum network throughput can be achieved and maintained using RL with fair bandwidth share for the nodes. Next, it shows how a stateless variant of RL can be used for collision-free transmission slot scheduling without network time synchronization. This is done using a slot defragmentation operation embedded with MAB components to reduce bandwidth redundancy arising from slot allocation in the absence of network time synchronization. To be noted, the framework proposed in this work is decentralized in the sense that all nodes learn the transmission schedule independently using localized network information.

### III. NETWORK AND TRAFFIC MODEL

The network models considered in this paper are generalized multi-point to point with arbitrary mesh topologies (Fig. 1) and traffic patterns. In order to understand and analyze the impacts of network information availability, both fully and partially connected topologies are considered. For fully connected, each node can possess complete network-wide information including congestion, throughput etc. For partially connected, a node can possess only localized information within its neighborhood.

As for packet generation, constant packet rate and Poisson distributed packets have been used. The MAC layer traffic load model is created such that a packet generated from a node is sent to one of its uniformly randomly chosen 1-hop neighbors. This is done on a packet-by-packet basis.

Networks without and with time slotting are investigated. In both cases, no network time synchronization is assumed. As described later in Section V, the network model includes the ability of piggybacking very low data-rate control information using parts of the data packets. Such control information is used for local information sharing needed by the RL learning.

### IV. REINFORCEMENT LEARNING AND MULTI-ARMED BANDIT

Reinforcement Learning (RL) is a model-free approach used to solve a Markov Decision Process (MDP) [15]. One of the commonly used RL techniques is a value-based tabular update method known as Q-Learning. Each entry in the table $Q(s, a)$ is a $Q$-value representing the importance of taking an action $a$ when the system is in state $s$. This table is updated by taking repeated actions stochastically with a bias towards the action with the highest $Q$-value, which is updated based on the acquired reward. For a received reward, the $Q$-value for a state-action pair is updated using the Bellman's temporal difference equation [15]. A special class of RL problems for non-associative settings are known as Multi-Armed Bandits (MAB), where there is no state abstraction and the agent's goal is to determine the best set of actions that would maximize its expected reward [15].

A variant of Q-table updates, used in multi-agent RL environments, known as Hysteretic Updates [15], is used in this work. Without knowing the actions taken by the rest of the agents, each agent learns to achieve a coherent joint behavior by observing the effects of its own actions on the system. The key challenge is that an agent's cumulative reward not only depends on its own actions, but also those of the others. Even if an agent takes a good action, it may still receive a penalty because of other agents' poor actions. Hysteretic Learning addresses this by assigning less importance to penalties as compared to the rewards by using two different learning rates. The higher learning rate is used if an agent's action produces desired beneficial effects. Otherwise, the lower learning rate is used so that lesser importance is given to actions that are deemed suitable by the agent but did not produce beneficial results probably due to unfavorable actions taken by the other agents in the environment. This prevents the Q-values of good actions to go down, thus accelerating learning convergence. A detailed description of RL, MAB and Hysteretic Learning can be found in [15] and [19].

### V. REINFORCEMENT LEARNING FOR RANDOM ACCESS MAC

#### A. Modeling Network Protocol Synthesis as MDP

Each network node acts as an independent RL agent and the wireless network acts as the environment through which the agents interact via their actions. In what follows, it is shown as to how node transmission behavior can be modeled as a Markov Decision Process (MDP), and when the MDP is solved using RL, it can give rise to probabilistic transmission strategies that represent a MAC protocol. The details of different RL components are as follows.

Actions: An RL agent's (i.e., a node) actions are represented by transmission probabilities in the range [0, 1]. Meaning, the action defined by the probability $p$ represents a packet transmission with that probability. The probabilities are discretized at equal intervals in order to keep the action space discrete. The interval size determines the action space size, and the resulting RL performance and convergence properties. In this work, the interval size of 0.05 is chosen empirically based on the performance and convergence speed tradeoffs. The learning error, represented as the difference between the throughput obtained via RL and that of a known benchmark, as described in the next subsection goes down, and convergence time goes up with increase in the size of the action space. The actions are selected following an $\epsilon$-greedy exploration policy, where the agents explore all the possible actions randomly with a probability $\epsilon$, and take the action based on the maximum Q-value with probability $1 - \epsilon$.

States: The state experienced by an agent/node is represented by the congestion level it encounters. A node estimates its state during a learning epoch from the number of packet collisions it experiences during the epoch. It is encoded as the collision probability computed as the ratio of number of collided to transmitted packets. As done for the action space, collision probabilities are also discretized into a fixed interval size (in range [0, 1]), which determines the state space size. There exists a tradeoff between learning performance and convergence time for different state space sizes. A state space size of 5 has been

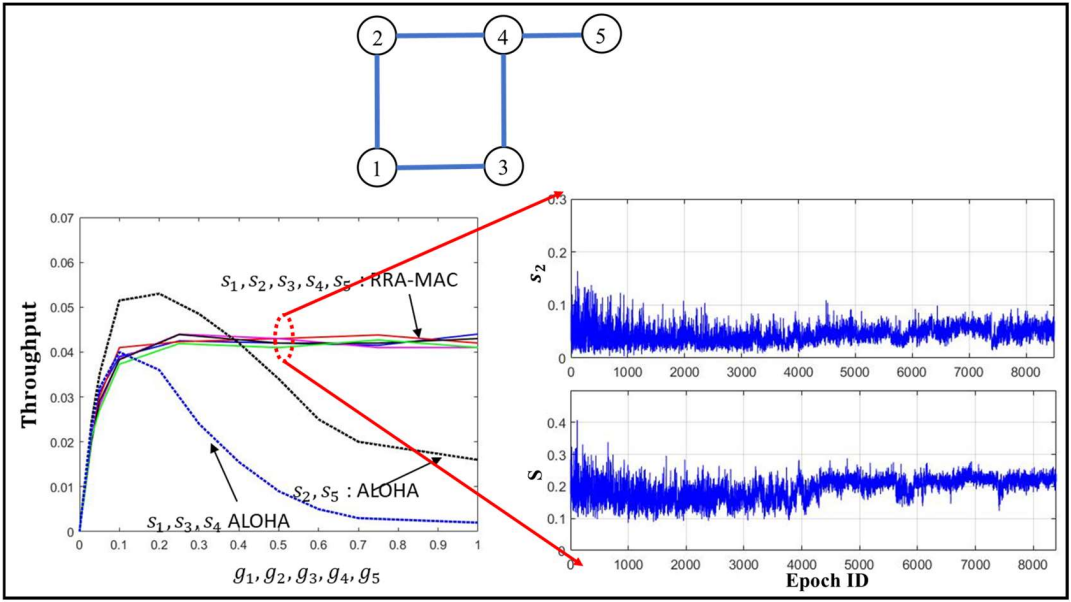
Fig. 2: RRA-MAC in a 5-node partially-connected topology and the learning convergence behavior

chosen empirically for all presented results in the this paper.

Reward: Since learning is node-independent and the nodes do not possess network-wide information, the reward is decided based on a node's localized information collected in-band using piggybacking over the MAC layer PDUs.

Let $s_i$ be the current throughput of node $i$ and $s_{i \to j}$ be the portion of node $i$'s throughput for which $j$ (one-hop neighbor of $i$) is the intended receiver. Node $j$ periodically piggybacks $s_{i \to j}$ in its outgoing MAC layer PDUs. Node $i$ then calculates its own throughput $s_i = \sum_{\forall 1-\text{hop neighbor } j} s_{i \to j}$, which it periodically piggybacks along with its one-hop neighbors' throughput $s_j$ in its outgoing PDUs.

Now, given that a node $i$ knows its own throughput as well as its two-hop neighbors' throughput (i.e., $s_i, s_j$), it calculates its localized neighborhood throughput as $S_i = s_i + \sum_{\forall j} s_j$. The packet transmissions from nodes that are within a 2-hop locality can lead to collisions at the receiver. Thus, throughput of a node is affected by its all 2-hop neighborhood transmission policies and hence, 2-hop neighborhood throughput is considered for reward formulation. Using this information, a reward function is formulated with the aim of maximizing network throughput while minimizing the deviation of throughputs of each individual node. Thus, an action is rewarded if both the throughput and fairness gradients as defined by $\Delta S_i = S_i(t) - S_i(t-1)$ and $\Delta f_i = f_i(t) - f_i(t-1)$ respectively are positive. Here, $f_i$ is the fairness coefficient computed as:

$$f_i(t) = -\sum_{\forall k \neq i} |s_i(t) - s_k(t)|$$

$k \epsilon$ onehop neighborhood of $i$.

Thus, a temporal gradient-based reward is formulated as follows.

$$R_i(t) = \begin{cases} +50, & \Delta S_i - \epsilon_s > 0, \Delta f_i - \epsilon_f > 0 \\ -50, & \text{otherwise} \end{cases} \quad (1)$$

Here, coefficients $\epsilon_s$ and $\epsilon_f$ are used so that the agents don't get stuck in a near optimal solution. Experimentally chosen learning hyper-parameters are set to: Hysteretic Learning rates of 0.9 and 0.1, and a discount factor of 0.95. Using this reward arrangement, each node independently learns a probabilistic transmission strategy such that the network wide throughput is maximized while attempting to maintain node-level fair bandwidth distribution. This behavior gives rise to the proposed RL-based

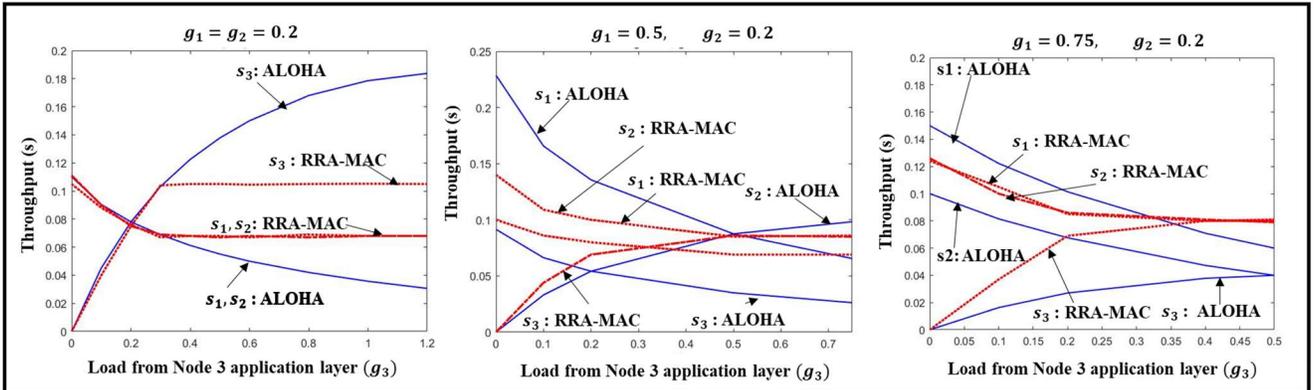
Fig. 3: Performance of RRA-MAC for heterogeneous loading conditions

Random Access MAC (RRA-MAC) Protocol. Note that although each node independently learns transmission policies, their learning process is mutually affected by the collisions caused by their individual actions. A learning convergence in such situation is when all nodes are able to choose the correct transmissions probabilities for given collisions in its up to 2-hop neighborhood.

### B. Results and Analysis

In this section we present the performance of RRA-MAC framework that uses RL to solve network protocol modeled as a Markov Decision Process (MDP). In Fig. 2, the performance of RRA-MAC is compared with the simplest known sensor/IoT random access, namely ALOHA, that does not rely on complex hardware features including carrier sensing and time-slotting. The figure shows performance for a 5-nodes partially-connected topology in which nodes 1, 2, 3 and 4 form a square and node 5 is connected only to node 4. The first observation is that unlike for ALOHA, RRA-MAC is able to provide a fair bandwidth distribution for all five nodes. Since nodes 1, 3 and 5 are topologically disadvantageous in that they experience higher collision rates compared to nodes 2 and 4, with ALOHA those three nodes experience lower overall throughputs. Such unfair access performance aggravates as traffic loading increases. The RL-based RRA-MAC circumvents that by using a fairness-aware reward structure. This allows the proposed learning-based mechanism to handle topological heterogeneity in a fair manner.

The second notable observation is that unlike the ALOHA family of protocols, the learning-based access can sustain high throughput at high loading conditions. With ALOHA, excessive collisions bring sustainable throughput down beyond a critical loading point. With RRA-MAC, this is avoided by the RL agents via learning to reduce transmission probabilities (i.e., actions) in states that indicate increasing collisions in the neighborhood. This causes the RRA-MAC throughput to be sustained at higher loads, while maintaining node level throughput fairness.

Fig. 2 also shows the learning convergence behavior for both network-wide and individual throughputs for individual node load $g_i = 0.5, 1 \leq i \leq 5$. Here, $g_i$ is the application layer load (Erlangs) in node $i$. Post convergence, the nodes learn to take actions so that network throughput ($S$) is maximized while maintaining fairness in available bandwidth distribution.

Performance of RRA-MAC in a 3-nodes fully-connected topology for heterogeneous traffic is shown in Fig. 3. With ALOHA access, there is a high variation of throughputs among the three nodes for heterogeneous load distribution. In contrast, with RRA-MAC, the differences in throughputs of individual nodes are significantly smaller. In each of the three plots in Fig. 3, the loads from node-1 ($g_1$) and node-2 ($g_2$) are kept fixed at different values, and the node-level throughput variations are observed for varying load from node-3 ($g_3$). These represent the scenarios: $g_1 \leq \hat{g}, g_2 \leq \hat{g}$, $g_1 \leq \hat{g}, g_2 > \hat{g}$ or $g_1 > \hat{g}, g_2 \leq \hat{g}$, and $g_1 > \hat{g}, g_2 > \hat{g}$. It can be observed that with DRLI-MAC, the RL agents in nodes learn to adjust the transmit probability such that the available wireless bandwidth is fairly distributed. Also notable is the fact that the RRA-MAC logic can hold the maximum fair throughput for higher network loads, even under heterogeneous loading conditions.

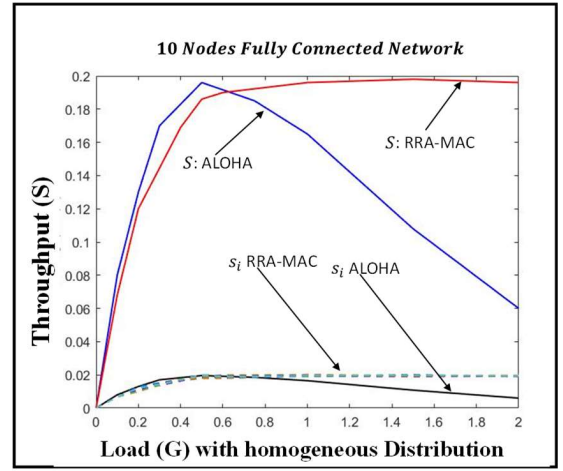

Fig. 4: Performance of RRA-MAC in fully-connected topology

The ability of the proposed mechanism to maximize and sustain network throughput in a fair manner for fully-connected topologies is shown in Fig. 4. Throughput attained using RRA-MAC increases, reaches a maximum and then sustains with increase in network load. Fig. 5 shows the ability of the RL-based protocol to adjust to dynamic network conditions. The ability to adapt to time-varying network traffic is shown for a 12-node partially connected topology in Fig. 5 (a). It can be observed that learning adjustment to a change in network load is faster as compared to the initial convergence. It is because, once a Q-table is learnt, the updated table maintains some information regarding which actions are better at a particular state representing a certain collision probability. Hence, the learning agent already has certain level of intelligence regarding the best sets of possible actions which helps it to converge quicker as compared to the case of fresh random initialization of Q-values. This effect can be further investigated on a dynamic node failure/node addition scenario as shown in Figs. 5 (b) and (c). While for the node failure scenario, convergence is faster than that of fresh start due to the reasons explained above, convergence does not speed up as much for the node additions. This is because, on addition of a node, it has to start its learning from the scratch with random initialization of Q-table, thus delaying the convergence.

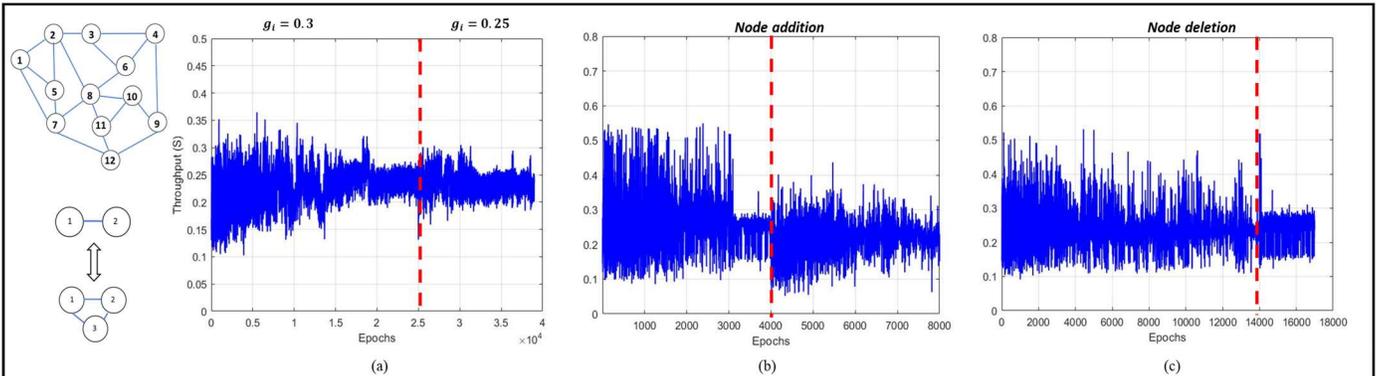

Fig. 5: Adjustment to dynamic network conditions by RRA-MAC

Effects of Channel Unreliability: To understand the robustness of the learning-based RRA-MAC to channel errors, performance was analyzed for different packet error probabilities. For a 3-nodes fully connected topology, throughput ratio ($\frac{S_{RRA-MAC}}{S_{ALOHA}}$) and convergence time were observed to be 1.83, 1.86, 1.82 and 6.5, 6.8 and 7.1 ($\times 10^3$ epochs) for packet error probability values of 0%, 5% and 10% respectively. With increase in packet error probability, a greater number of packets gets dropped. This makes each node to require more learning epochs to get an estimate of the correct neighborhood throughput to compute rewards and to update the Q-table values. Although the convergence slows with more channel errors in general, the slowdown is acceptable for the practical range of packet error probabilities $(0-10\%)$. Similarly, the post-convergence throughput ratio remains in the same ballpark value for different values of packet error probabilities up to 10%. This indicates that the impacts of channel errors on RRA-MAC are no worse than those on the ALOHA protocol logic.

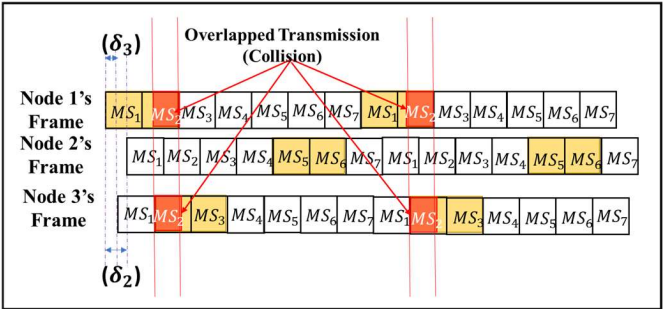

Fig. 6. Asynchronous frames in a 3-nodes fully-connected network

To summarize, Reinforcement Learning for medium access in wireless network can make nodes learn transmission policies in a cooperative manner in order to maximize throughput and fairness. This is achieved in a resource constrained system in the absence of complex hardware support such as time-slotting, carrier-sensing, time-synchronization, and collision detection, thus making it suitable for low-complexity IoT and sensor nodes.

## VI. MULTI-ARMED BANDIT LEARNING FOR TIME-ASYNCHRONOUS TDMA MAC

In the presence of time slotting, MAC packet collisions can be largely avoided by TDMA-based packet transmission scheduling. This section presents a learning mechanism towards that goal, specifically when network time-synchronization is not available. High resolution and accurate time-synchronization over wireless can be expensive, especially in low-cost sensor and IoT nodes with limited processing and communication resources. Moreover, performance of TDMA MAC protocols that rely on network time-synchronization can be very sensitive to time-synchronization drifts. This section shows how MAC layer packet scheduling can be learned in the absence of time-synchronization using Multi-arm Bandit (MAB) techniques.

Since time is not synchronized, the scope of a node's TDMA frame is strictly local. It decides the start time of its own frame, and the end time is decided based on a predefined frame duration, which is denoted by $T_{frame}$. The node does not know about the start times of the other nodes' frames. Within a frame, a node can schedule a packet transmission only in certain discrete time instances away from its frame start time. The intervals between those time instances are referred to as mini-slots, the duration of which is an integer submultiple of the fixed size packet duration, and is equal at all nodes.

This arrangement of mini-slot-based asynchronous TDMA is shown for a 3-node fully connected network in Fig. 6. Frames of nodes 2 and 3 lag from that of 1 by $\delta_2$ and $\delta_3$ durations. Here, the frame size equals 7 mini-slots and a mini-slot is half of packet duration. A node can select any of these 7 mini-slots within its frame as the starting point of its packet transmission. The figure depicts a situation where for packet transmissions, nodes 1, 2 and 3 select mini-slots 1, 5 and 2 respectively in their own frames and periodically transmit in those mini-slots in subsequent frames. Packets from nodes 1 and 3 collide because of their time-overlapped transmissions (indicated by red), whereas packets from 2 are successfully transmitted.

The transmission scheduling problem in this context boils down for each node to be able to choose a start-transmission mini-slot within its own frame, and that is without colliding with other nodes. Such collision-free mini-slots should be selected locally at each node in a fully independent manner without the help of any centralized allocation coordinators and network time-synchronization. This is achieved by the framework comprising of two distinct components: MAB-learning-based slot (mini-slot) scheduling and slot-defragmentation operation to minimize any bandwidth redundancy resulting from the time-asynchronous scheduling by MAB. The entire flow is captured in Fig. 7.

This slot allocation problem can be modeled as a multi-Agent MAB. Each node in this scenario acts as an independent '$f$-armed bandit', where $f$ is the frame size in number of mini-slots. In other words, the action of an agent is to select a start-transmission mini-slot in the frame. The MAB environment is the wireless network itself through which the bandits interact via the selection of the arms (i.e., start-transmission mini-slots). The reward is designed such that the bandit receives a reward of $+1$ if the packet transmission in the selected mini-slot is successful. Else, a penalty of $-1$ is assigned.

Using this MAB model, all nodes individually learn collision-free transmission schedules in an independent manner. Fig 7 (Stage 1) shows the learning convergence in a 3-node fully-connected network for a constant data rate $\lambda = 1$ packet per frame per node, and the number of arms $f = 4$. Packet transmission dynamics by all nodes are plotted in the figure with node 1's frame as the frame of reference. Frames of nodes 2 and 3 lag that of node 1 by $0.4\,\tau$ and $0.75\,\tau$ respectively, where $\tau$ represents the packet duration. Note that while there are collisions initially, after learning convergence, the nodes learn to select collision-free start-transmission mini-slots. Such learning takes place without network time-synchronization.

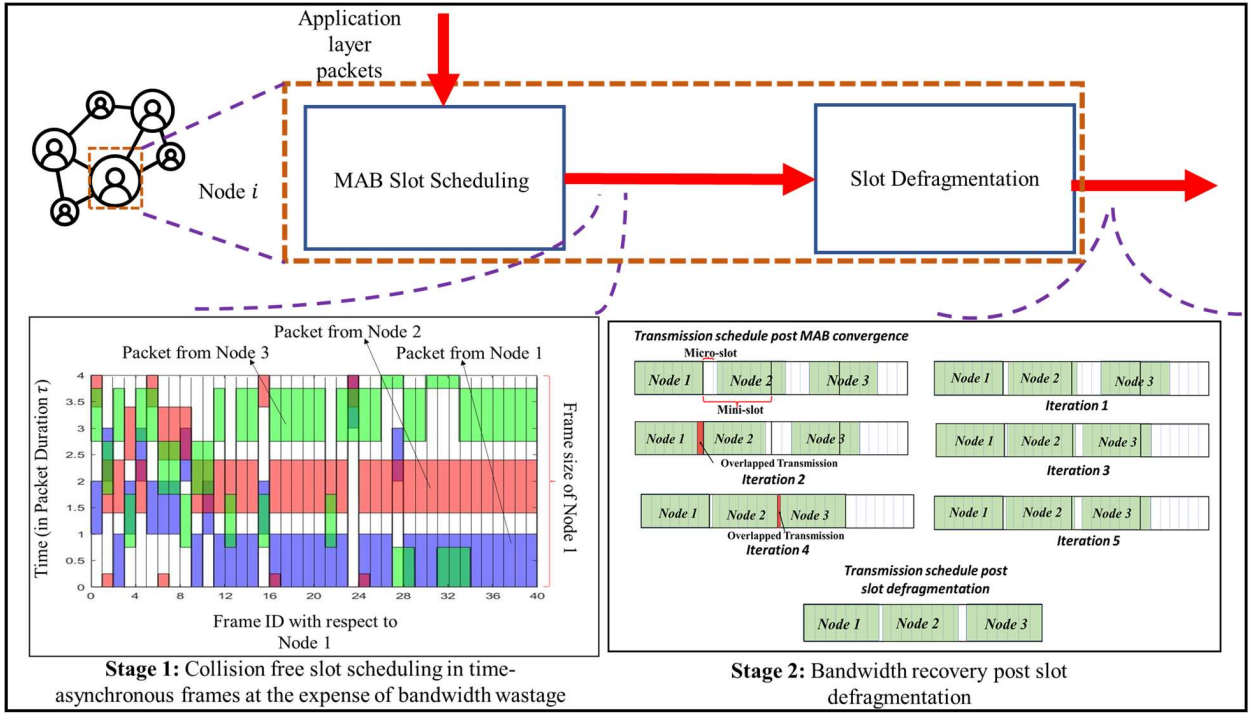

Fig. 7. MAC slot scheduling in time-asynchronous networks using Multi-arm Bandit learning

Since this framework requires each node to perform its own iterative search for a collision free start-transmission mini-slot independently, short-term collisions and scheduling deadlocks can occur. This can be mitigated by making frame size $f$ larger than the absolute needed minimum $f_{min}$ in the presence of time-synchronization. This leads to certain amount of bandwidth redundancy and is represented by a factor $K$, defined as $K = \frac{f}{f_{min}}$.

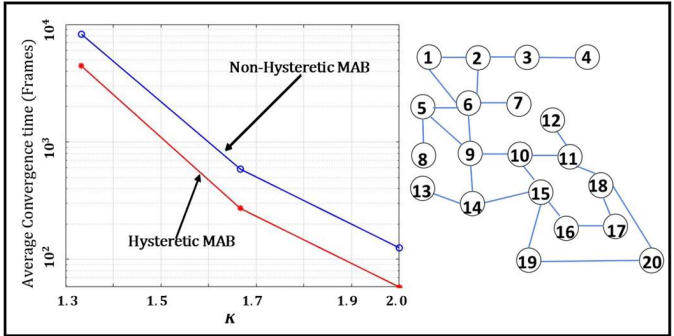

Fig. 8: Convergence time variation with $K$

This bandwidth redundancy factor plays a significant role in the MAB learning convergence speed. This can be observed from Fig. 8, which shows that for a 20-node mesh network, learning convergence speeds up with larger $K$. It is because with increase in $K$, the number of feasible solutions of the MAB problem increases and hence the probability of finding a collision-free transmission strategy increases. Also, the convergence speed is observed to be high with Hysteretic learning as compared with the classical MAB update rule [11, 16]. This is achieved by giving less importance to penalties than rewards in Hysteretic MAB as explained in Section IV.

As observed in Fig 8, convergence of MAB learning speeds up with increased bandwidth redundancy factor $K$. However, increased $K$ leads to an increase in frame length which in turn increases bandwidth wastage. This redundancy can be mitigated by the following *slot defragmentation* mechanism after the MAB learning converges.

Slot defragmentation is implemented by discretizing each mini-slot within a frame into '$s$' micro-slots. After MAB convergence, each node shifts its transmission by one micro-slot back in time till it experiences a collision. Upon experiencing a collision, the node undoes its previous shift action to find a new transmission micro-slot. In this way, the nodes estimate the unused space in the frame and try to reduce it in a coordinated manner. The logic for defragmented backshift executed by each node $i$ is given in Algorithm 1.

This mechanism of defragmentation for a 3-node fully-connected network is shown in Fig. 4 (Stage 2). It shows how the frame structure (with respect to node 1) evolves over 5 iterations of the defragmentation process for bandwidth redundancy factor $K = 1.33$ and 7 micro-slots ($s = 7$). Node 1 does not shift its transmission since it is transmitting at the beginning of the frame. Nodes 2 and 3 backshift their transmissions by one micro-slot per iteration. In iteration 2, nodes 1 and 2 experience collision. Hence node 2 undoes its previous action by shifting by one-micro-slot forward in iteration 3. But node 1 does nothing in iteration 3 since it experienced a collision without any micro-slot shift in its previous frame. Similarly, nodes 2 and 3's packets collide in iteration 4 because of backshift operation of node 3. Node 3 shifts forward its transmission by one micro-slot and knows that it has found its suitable transmission micro-slot. In this example, the new frame size as shown in the figure reduces by 21% because of slot defragmentation. This bandwidth redundancy left after slot defragmentation is due to the time lag existing among the nodes resulting from the lack of network synchronization.

```
1:  Initialize: μ_shift_i = 0, c_i = 0    // μ_shift_i: Number of micro-slot
    that node i has shifted; c_i: Status of the micro-slot search (1, if
    search is complete, else, 0)
2:  If (! Tx in the beginning of frame), do:
3:      Shift to previous micro-slot
4:      μ_shift_i ++
5:      Check Collision
6:      If (Collision ==TRUE):
7:          Check action in the previous frame a(t−1)
8:          If (μ_i(t) > μ_i(t − 1)):
9:              Shift to next micro-slot
10:             Check Collision
11:             If (Collision ==TRUE):
12:                 Shift to previous micro-slot
13:             End If
14:         Else If (μ_i(t) < μ_i(t−1)):
15:             Shift to next micro-slot
16:         End If
17:         Set c_i = 1
18:         Piggyback c_i, μ_shift_i
19:         Check c_j, ∀j ∈ one − hop neighbor
20:         If (c_j == 1 (∀j ∈ one−hop neighbor))
21:             Find new frame size:
22:             F_shrunk(t) = max{μ_shift_i(t), μ_shift_j(t)}
23:             If (F_shrunk(t) == F_shrunk(t − 1)):
24:                 Frame Size ← Frame Size − F_shrun
25:                 μ(t) = μ(t − 1) − F_shrun
26:                 Ignore all collisions
27:             End If
28:     Else:
29:         Do Nothing
30:         Set c_i = 1
31:         Piggyback c_i, μ_shift_i
32:         Check the value of c_j, ∀j ∈ one−hop neighbor
33:         If c_j == 1 (∀j ∈ one−hop neighbor)
34:             Find new frame size:
35:             F_shrun(t) = max{μ_shift_i(t), μ_shift_j(t)},
36:             Frame Size ← Frame Size − F_shrunk
37:             μ(t) = μ(t − 1) − F_shru
38:         End If
39: End If
```

**Algorithm. 1.** Defragmented Backshift

Once a node finds a stable micro-slot, it piggybacks over data packets the information about the number of micro-slots it has shifted ($\mu\_shift$) to find its final stable position. Thus, a node $i$ knows that its one-hop neighbors have found stable micro-slots. It then computes the new frame size by subtracting the maximum of the $\mu\_shift$ values (i.e., received from its neighbors) from its current frame size.

Upon performing slot defragmentation in a 9-node fully-connected topology with the bandwidth redundancy factor $K$ set at 1.67, the bandwidth redundancy goes down from 67% to 3.3%. The bandwidth redundancy of 3.3% at the end of defragmentation is caused primarily by the temporal lags across the frame start times. Similarly, for a partially connected topology shown in Fig. 9, for $K = 2$, bandwidth redundancy after defragmentation reduced from 100% to 7.11%.

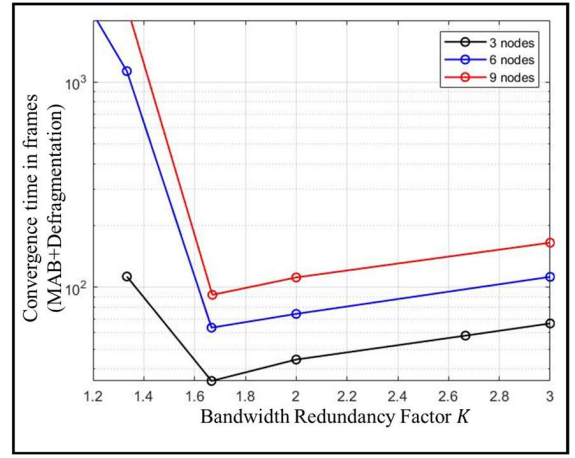

Fig. 9: Convergence time variation with $K$ for fully connected networks

Fig. 9 depicts the additive time for stage-1 MAB convergence and stage-2 defragmentation convergence. Larger $K$ values speed up MAB convergence while slowing down the defragmentation process. The latter is because with a larger frame length, the number of iterations that a node has to backshift its transmission micro-slot to find a suitable micro-slot increases. Thus, the search space to find the suitable transmit micro-slot increases with $K$. As can be seen in Fig. 10, the total convergence duration (MAB and slot defragmentation) initially goes down with increase in $K$, reaches a minimum, and then goes up again. This is because for small $K$, MAB convergence time is significantly higher than defragmentation convergence and hence the total convergence is largely affected by the MAB learning convergence. However, for larger $K$ values, defragmentation convergence time overpowers MAB convergence time, and thus, total convergence time increases with $K$. These results indicate that an optimum value of $K$ exists that gives the minimum total convergence time of the proposed learning framework.

VII. SUMMARY AND CONCLUSIONS

The concept of network protocol synthesis using RL and MAB is explored in this article. Here, Reinforcement Learning (RL) and Multi-Armed Bandits (MAB)-based approaches for wireless network protocol synthesis are summarized and a comprehensive distributed RL and MAB-based framework is presented that can synthesize MAC protocols for both random access and time-slotted systems which can overcome the drawbacks of the existing approaches. One notable feature of the framework is that it does not rely on complex hardware features such as collision detection, time synchronization, and carrier sensing, thus making it suitable for ultra-resource constrained sensor and IoT nodes. The learning-based framework allows nodes to learn in an independent manner to maximize network throughput and to maintain fair bandwidth distribution, even in heterogeneous network topologies and loading conditions. It is also shown how the developed mechanism makes the IoT nodes learn transmission scheduling policies to avoid collisions in a time-slotted system without network time-synchronization. Future work on this topic includes exploring other access performance parameters of the protocol, such as, end-to-end delay, energy efficiency etc. and generalizing the framework for protocol synthesis in networks with or without any resource-constraints.

## VIII. REFERENCES

1. Leon-Garcia, Alberto, and Indra Widjaja. Communication networks: fundamental concepts and key architectures. Vol. 2. New York: McGraw-Hill, 2000.
2. Lee, Jin-Shyan, Yu-Wei Su, and Chung-Chou Shen. "A comparative study of wireless protocols: Bluetooth, UWB, ZigBee, and Wi-Fi." IECON 2007-33rd Annual Conference of the IEEE Industrial Electronics Society. Ieee, 2007.
3. Chu, Yi, Paul D. Mitchell, and David Grace. "ALOHA and q-learning based medium access control for wireless sensor networks." ISWCS, IEEE, 2012.
4. Z. Lan, H. Jiang and X. Wu, "Decentralized cognitive MAC protocol design based on POMDP and Q-Learning," 7th International Conference on Communications and Networking, China, 2012, pp. 548-551.
5. Ali, Rashid, et al. "Deep reinforcement learning paradigm for performance optimization of channel observation–based MAC protocols in dense WLANs." IEEE Access 7 (2018): 3500-3511.
6. Yu, Yiding, Taotao Wang, and Soung Chang Liew. "Deep-reinforcement learning multiple access for heterogeneous wireless networks." IEEE Journal on Selected Areas in Communications 37.6 (2019): 1277-1290.
7. Lee, Taegyeom, and Ohyun Jo Shin. "CoRL: Collaborative Reinforcement Learning-Based MAC Protocol for IoT Networks." Electronics 9.1 (2020): 143.
8. Galzarano S., Liotta A., Fortino G., "QL-MAC: A Q-Learning Based MAC for Wireless Sensor Networks", Algorithms and Architectures for Parallel Processing, 2013, Lecture Notes in Computer Science, vol 8286. Springer.
9. Savaglio, Claudio, et al. "Lightweight reinforcement learning for energy efficient communications in wireless sensor networks." IEEE Access 7 (2019): 29355-29364.
10. Bayat-Yeganeh, Hossein, Vahid Shah-Mansouri, and Hamed Kebriaei. "A multi-state Q-learning based CSMA MAC protocol for wireless networks." Wireless Networks 24.4 (2018): 1251-1264.
11. Park, Huiung, et al. "Multi-agent reinforcement-learning-based time-slotted channel hopping medium access control scheduling scheme." IEEE Access 8 (2020): 139727-139736.
12. Liu, Jiahao, et al. "Dynamic channel allocation for satellite internet of things via deep reinforcement learning." 2020 International Conference on Information Networking (ICOIN). IEEE, 2020
13. Ahmed, Faisal, and Ho-Shin Cho. "A time-slotted data gathering medium access control protocol using Q-learning for underwater acoustic sensor networks." IEEE Access 9 (2021): 48742-48752.
14. Park, Sung Hyun, Paul Daniel Mitchell, and David Grace. "Performance of the ALOHA-Q MAC protocol for underwater acoustic networks." 2018 International Conference on Computing, Electronics & Communications Engineering, IEEE.
15. Sutton, Richard S., and Andrew G. Barto. Reinforcement learning: An introduction, MIT press, 2018.
16. H. Dutta and S. Biswas, "Towards Multi-agent Reinforcement Learning for Wireless Network Protocol Synthesis," 2021 International Conference on COMmunication Systems & NETworkS (COMSNETS), 2021, pp. 614-622, doi: 10.1109/COMSNETS51098.2021.
17. H. Dutta and S. Biswas, "Medium Access using Distributed Reinforcement Learning for IoTs with Low-Complexity Wireless Transceivers," 2021 IEEE 7th World Forum on Internet of Things (WF-IoT), 2021, pp. 356-361, doi: 10.1109/WF-IoT51360.2021.9595062.
18. Dutta, Hrishikesh, and Subir Biswas. "Distributed Reinforcement Learning for scalable wireless medium access in IoTs and sensor networks." Computer Networks 202 (2022): 108662.
19. Matignon, Laëtitia, Guillaume J. Laurent, and Nadine Le Fort-Piat. "Hysteretic q-learning: an algorithm for decentralized reinforcement learning in cooperative multi-agent teams." 2007 IEEE/RSJ International Conference on Intelligent Robots and Systems.